# Statistics for the Dynamic Analysis of Scientometric Data:

# The evolution of the sciences in terms of trajectories and regimes

Loet Leydesdorff

Amsterdam School of Communication Research (ASCoR)

Kloveniersburgwal 48, 1012 CX Amsterdam, The Netherlands;

loet@leydesdorff.net ; http://www.leydesdorff.net

**Abstract**

The gap in statistics between multi-variate and time-series analysis can be bridged by using entropy statistics and recent developments in multi-dimensional scaling. For explaining the evolution of the sciences as non-linear dynamics, the configurations among variables can be important in addition to the statistics of individual variables and trend lines. Animations enable us to combine multiple perspectives (based on configurations of variables) and to visualize path-dependencies in terms of trajectories and regimes. Path-dependent transitions and systems formation can be tested using entropy statistics.

**Introduction**

The sciences evolve as complex and non-linear systems. Such systems can be expected to contain recursive and interactive terms. Recursion means that the previous state of the system (at *t*-1) is potentially a main driver for its further development, and interactions can occur, for example, in university-industry-government relations among differently codified subsystems of communication (Leydesdorff & Zawdie, 2010; cf. Luhmann, 1995). Within the sciences under study in scientometrics, specialties and disciplines can be considered as discursive subsystems with their own dynamics; interactions and translations can lead to "interdisciplinarity," "transformative science," and/or newly emerging developments (Leydesdorff, Cozzens, & Van den Besselaar, 1994).

Complex systems of scientific communications can be expected to drive themselves towards the edge of chaos (e.g., Kauffman, 1993; Prigogine & Stengers, 1984). In addition to large-scale crises and paradigm changes—which may be rare—one can expect smaller-scale developments to reorganize the sciences continuously (Bak *et al.*, 1987; cf. Bak & Chen, 1991). The sciences as volatile systems of rationalized expectations gain stability because of textual organization and archival aggregation, community formation and institutionalization. Scientists follow the research fronts reflexively and organize the expanding horizons of possibilities in historical contexts. Validated knowledge, for example, is more stable than knowledge claims in manuscripts.

In my opinion, multivariate analysis in scientometrics has hitherto focused mainly on static designs that may be of limited value when addressing questions about nonlinear developments. Significance testing and boot-strapping techniques, for example, have been developed using available software such as SPSS, SAS, Stata, or more recently also R. New programs of social network analysis and visualization have added new statistics such as centrality measures, and mechanisms such as preferential attachment have been theorized (Barabási & Link, 1999). The specification of a mechanism, however, is based on a dynamic assumption (Price, 1976).

In general, evolutionary mechanisms cannot be observed directly—they are genotypical—but the phenotypical footprints of the mechanisms can be analyzed (Hodgson & Knudsen, 2011; Luhmann, 1990). For example, the highly skewed distributions which prevail in the study of the sciences (Seglen, 1992) can be considered as the result of strong (recursive) selections and possibly hyper-selection that operates on the variation. Using cross-lagged correlation analysis, Cook & Campbell (1979) "quasi-experimented" with systems containing two subdynamics (e.g., "variation" and "selection"). Most of the curve-fitting and extrapolation in time-series analysis, however, focuses on single variables. Interactions among evolving subsystems cannot be tested using ARIMA (Auto-regressive Integrated Moving Averages) or similar such models in SPSS.[1]

Many researchers reduce dynamic problems to comparatively static ones. One measures in a specific year and similarly in the next year, and then one assumes that the difference can be considered as "the evolution" of the system under study. For example, publication and/or citation rates for different countries are plotted against time, and then one can assess whether the "USA is losing ground in science" or "China is emerging" with a linear or exponential dynamics by curve-fitting the individual trendlines. However, this does not tell us much about the dynamics of the publication system; it teaches us something about the relative contributions of subsystems to the system. A (so-called Markov) systemic assumption could, for example, be that the present state of the system is the best prediction of its next state. Deviations at $t+1$ from this present state ($t$) can perhaps be tested for their significance.

---

[1] More recently, Vector Autoregression Models are developed that capture linear interdependencies among a limited number of time series (Abdulnasser, 2004).

**The problem of the dynamics of science**

If one assumes that the sciences develop as socio-cognitive systems, the scientometrician has to assume that not only are the observable social or textual variables important, but also their (potentially hybrid; Callon *et al*., 1986) organization in terms of groups, clusters, factors, etc. The groupings, however, remain theoretical constructs. When both the observable variables and the structural variables (e.g., eigenvectors) can change over time, one obtains a system of partial differential equations that can neither be solved analytically nor tested unambiguously in terms of the development of the data.

In a systems-dynamical model one can try to estimate parameters using data. However, the data is then used restrictively as a test-bed for specific assumptions about the dynamics, whereas the scientometrician (and the statistician) may be interested in exhausting the information contained in the data. How can this rich information be used to inform and induce—in heuristic terms—our theorizing about the dynamics in the science (s) under study? Let me propose two venues that have been explored hitherto: entropy statistics and a dynamic extension of multidimensional scaling (MDS). These two venues share a non-parametric orientation, but were otherwise developed in different domains and in different periods of time.

*a. Entropy statistics*

Not only do word frequencies or citation frequencies change over time, but the words and references may also change in meaning because the (latent) structures in the data evolve at the same time (Leydesdorff, 1997). Structural elements may emerge, bifurcate or merge in processes of fusion and fission in the data under study (Callon *et al*., 1991). Parameter estimation in a structural equation model, however, is of limited value if the parameters can be expected to change with the values of the variables.

When I first ran into these problems studying the relations of words and co-words in full texts in terms of their cognitive structures, Theil's (1972) "statistical decomposition analysis" provided a data-oriented approach to combining the static and dynamic analyses. Based on Shannon's (1948) information theory, decomposition analysis offers a non-parametric statistics in terms of bits of information at each moment in time (probabilistic entropy), while changes can be quantified using Kullback & Leibler's (1951) divergence measure of information.

Let me briefly recapitulate the idea. First, the data can be considered as a relative frequency distribution in several dimensions. In addition to the frequency distribution, one can, for example, consider the factor structure or the next-order organizational level as a second dimension. A word can thus be assigned not only an occurrence frequency, but also a position in the sentence, paragraph, section, or full text.

Shannon's information measure allows for adding these dimensions as subscripts to the probability distribution. Shannon's formula:

$$H = -\sum_i p_i \log_2(p_i) \qquad (1)$$

can be extended as follows:

$$H = -\sum_i p_{ijk\ldots} \log_2(p_{ijk\ldots}) \qquad (2)$$

This multivariate extension is unlimited in terms of the number of dimensions (*i*, *j*, *k*, …, etc).

Using the example of words in contexts, one can also consider one of the variables as a grouping versus the other as grouped variables. Theil (1972, at pp. 20f.) derived the decomposition algorithm as follows:

$$H = H_0 + \sum_G P_G H_G \qquad (3)$$

In this formula, $H_G$ denotes the uncertainty that prevails at the level of each group (e.g., class), $P_G$ is the relative frequency of that group in the sample, and $H_0$ is the between-group uncertainty. This algorithm can be nested, so that a word can be considered as part of a sentence, whereas the sentence is part of a paragraph, section, full text, etc. (Leydesdorff, 1991). In Leydesdorff (1995), this was elaborated both statically and dynamically for a set of 18 (full-text) articles in biochemistry.

The relation to the dynamic analysis is provided by the dynamic equivalents of these formulas. Given *a priori* uncertainty in a (multivariate) distribution, *a posteriori* uncertainty implies a (necessarily positive;[2] cf. Theil, 1972, at pp. 59f.) change that can be expressed as a Kullback-Leibler divergence as follows:

$$I_{q|p} = \sum_i q_i \log_2(q_i / p_i) \qquad (4)$$

In this formula, $\sum_i p_i$ denotes the *a priori* distribution and $\sum_i q_i$ the *a posteriori* one. The extension to the multivariate case is equally valid, and the extension to the decomposition straightforward (although more complex). One advantage is that all changes can be measured quantitatively in bits of information and decomposed in terms of contributing elements (cf. Krippendorff, 1986).

A further extension of these Shannon-based measures is provided by mutual information between two and among more than two variables. Mutual information among three dimensions can no longer be considered a Shannon entropy (Krippendorff, 2009), but is a *signed* information measure (Yeung, 2008, at pp. 59f.): it enables us to measure the difference between the redundancy generated in loops (spurious correlations) among the variables, and the uncertainty generated in their interactions. In another context, Leydesdorff (2003) developed this into a Triple-Helix indicator (Park & Leydesdorff, 2010; Lengyel & Leydesdorff, 2011; Strand & Leydesdorff, in press).

---

[2] Probabilistic entropy ($H$) is coupled to thermodynamic entropy ($S$) by Gibbs entropy formula: $S = k_B * H$. In this formula, $k_B$ is the Boltzmann constant with dimensionality Joule/Kelvin. $H$ is dimensionless and, for example, expressed in bits of information. Because of the second law both $S$ and $H$ necessarily increase or, in other words, the change in information (uncertainty) is always positive.

b. *Dynamic Multidimensional Scaling*

In information theory, assumptions about the shape of the distributions are not needed, since this calculus can be used with any relative frequency distribution (Bar-Hillel, 1955). However, the approach is data-oriented and tends to remain descriptive. Testing the results statistically requires the specification of an expectation and possibly assumptions about the shape of the distribution (for example, when one uses parametric statistics such as the analysis of variance). The different (engineering) background of the information-theoretical approach does not match easily with available statistics, which mostly find their origins in biology and psychology. However, alternative hypotheses can sometimes be tested against one another using information theory. In his almost 2000-pages *Handbook of Parametric and Nonparametric Statistical Procedures*, for example, Sheskin (2011) mentions information theory only in this context when discussing the likelihood ratio (at pp. 384 ff.). Using such an alternate approach, Leydesdorff (2000) showed that the European Union of that time (with 15 member states) did not develop into a single publication system, but should be considered as a set of national publication systems (cf. Leydesdorff & Oomes, 1999).

A breakthrough in combining static and dynamic analyses using standard statistics became possible more recently when Gansner *et al.* (2005) improved on the energy-minimizing algorithm for the mapping used by Kamada & Kawai (1989; cf. Ernten *et al.*, 2004). As with multidimensional scaling (MDS), the map can be considered as the result of minimizing a stress-function. MDS is part of the toolbox of methods in multivariate analysis that is available in the major software packages (Kruskall & Wish, 1973). Unlike factor and cluster analysis, MDS is oriented toward the *visualization* of structures in the data by reducing the dimensionality stepwise to two (or three) dimensions. This reduction in the dimensionality generates stress in the representation which can then be expressed as a statistics such as Kruskall's (1964) stress function.[3]

By minimizing not the stress itself, but the so-called *majorant* of the stress, both methodological and computational advantages can be realized. Baur & Schank (2008) proposed to combine minimization of the stress at each moment with minimization of the stress over time using the following formula:

$$S = \left[ \sum_t \sum_{i \neq j} \frac{1}{d_{ij,t}^2} (\|x_{i,t} - x_{j,t}\| - d_{ij,t})^2 \right] + \left[ \sum_{1 \leq t < |T|} \sum_i \omega \|x_{i,t} - x_{i,t+1}\|^2 \right] \qquad (4)$$

In Equation 4, the left-hand term is an equivalent to Kruskall's static stress function (Leydesdorff & Rafols, 2012), while the right-hand term adds the dynamic component, namely the stress between subsequent years. In this formula $\|x_i - x_j\|$ denotes the observable distance between two points on the map and $d_{ij}$ the distance in the underlying data matrix or, in other words, the multivariate space. The parameter $\omega$ can be considered as a weight that can be used to dampen or augment the visibility of change over time. This dynamic extension penalizes movements of the position of node *i* at time *t* ($\vec{x}_{i,t}$) toward its next position ($\vec{x}_{i,t+1}$) by increasing the stress value. Thus, stability is provided in order to preserve the mental map between consecutive layouts so that an observer can easily identify corresponding graph structures (Liu & Stasko, 2010; Misue *et al.*, 1995).

---

[3] The stand-alone program VOSViewer uses the approach of MDS, but the current version does not provide stress values (Van Eck & Waltman, 2010).

In other words, the configuration for each year is optimized in terms of the stress in relation to the solutions for previous years and in anticipation of the solutions for following years. The algorithm allows us (and its implementation in *Visone* enables us; cf. Baur *et al.*, 2002; Brandes & Wagner, 2004) to extend this to more than a single time-step. Note that the approach is different from those that take the solution for the previous moment in time as a starting position for iterative optimization. The nodes are not repositioned given a previous configuration; but the previous and the next configurations are included in the algorithmic analysis for each year. As against comparative statics, the (potentially non-linear) dynamics are animated (cf. Noyons & Van Raan, 1998).

Leydesdorff & Schank (2008) developed this idea as a network visualization for scientometric data using *Visone*, an existing program for network visualization and analysis (Brandes & Wagner, 2004). Using developments such as the emergence of nanoscience and nanotechnology as a separate discipline at the end of the 1990s, we could show that first the multidisciplinary journal *Science* and the year thereafter *Nanotechnology* functioned as catalyzing agents in a path-dependent transition (Figure 1). These journals became brokers between existing clusters of journals in applied physics and chemistry specialties, as can be indicated using their betweenness centrality in the relevant environments. After the transition an integrated structure encompassing these specialties emerged. (For the complete animation see http://www.leydesdorff.net/journals/nanotech.)

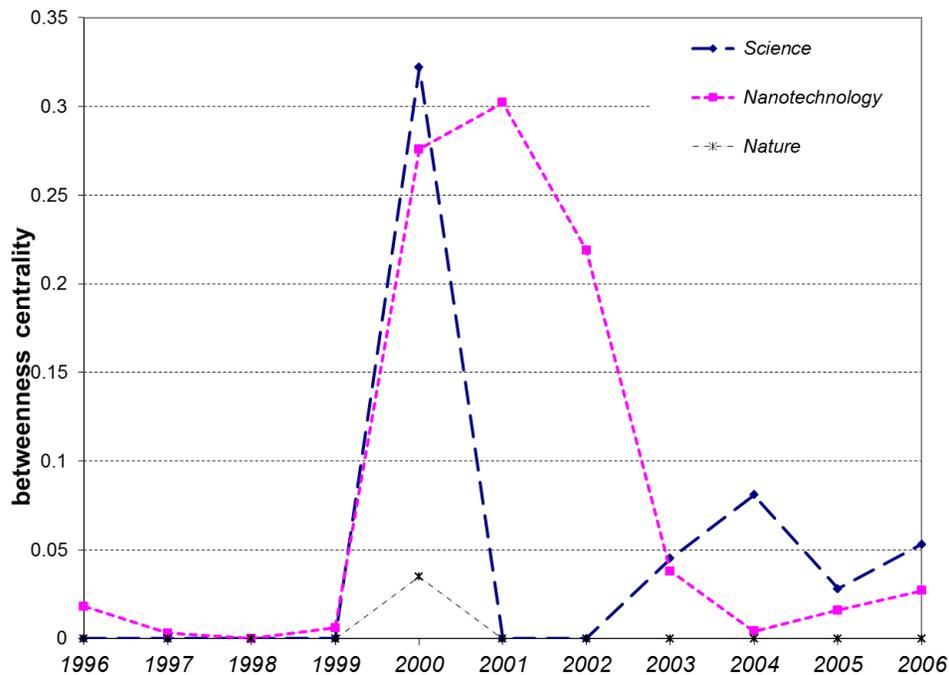

**Figure 1**: The development of betweenness centrality of the journals *Science* and *Nanotechnology* in the journal citation network during the path-dependent transition towards a nano-regime (1999-2003).

More recently, Leydesdorff (2011) elaborated this approach further by projecting the eigenvectors as constructs among the variables into the animations (e.g., at http://www.leydesdorff.net/eigenvectors/commstudies/). Thus, not only the evolution of observable variables can be made visible, but also the evolution of latent constructs. In principle, it would be possible to decompose the resulting stress into dynamic and static components.[4]

In the meantime, various programs have become available that allow for smoothing the interpolation of network representations at each moment when placed sequentially over time (e.g., PajekToSvgAnim, Gephi, Commetrix). It seems to me that a research program using the statistical approach is needed for the assessment of quality of (or error in) the animations of developments of data over time (Leydesdorff, Schank, Scharnhorst, & de Nooy, 2008).

**From multi-variate and multi-level statistics toward multiple perspectives**

A focus on multi-dimensionality in the representations and animations over time raises a more general question about the variables-based model that prevails in statistics. More important than individual variables are configurations of variables that generate perspectives on the non-linear dynamics. This requires a transition from working with a summary statistics (such as the mean) towards working with multivariate distributions.

Using information theory, uncertainty is no longer considered as error, but measured as an indicator (e.g., in bits). Theil's (1972) standard work about entropy statistics, for example, can be considered as a plea in economics that one should not compute with only the average pro-capita income, but also with the uncertainty contained in the income distribution. A distribution can first be considered linearly as a vector in a single dimension, but then be extended to a two-dimensional matrix, a three-dimensional cube of information, a four-dimensional hypercube, etc.

---

[4] The dynamic version of Visone is freely available at http://www.leydesdorff.net/visone.

A cube, for example, can be observed from three orthogonal perspectives given a geometrical space. Trajectories can be visualized in this space when distributions operate selectively upon each other in processes of "mutual shaping" (McLuhan, 1964). Using feedback, some selections can be selected for stabilization along a trajectory.

Furthermore, some stabilizations can be selected for globalization: three selection mechanisms operating upon one another—a triple helix—are sufficiently complex for modeling all species of chaotic behavior, such as hyperstabilization into a lock-in, path-dependencies through critical regions (Arthur, 1989; David, 1985; cf. Malerba *et al.*, 1999), and globalization as a possible consequence of meta-stabilization (Dolfsma & Leydesdorff, 2009). In other words, one can expect a difference when the system climbs up-hill along a trajectory under the regimes of full daylight or in the dark (Leydesdorff, 2002).

Using the science-push model of innovation, for example, one can assume that discoveries and inventions are lab-based, further upscaled along trajectories (cf. Nagin, 2005), and increasingly subject to the diffusion dynamics as a regime different from the production dynamics. Path-dependencies in these developments can be measured as critical transitions (using entropy statistics; Leydesdorff, 1991). Systems formation can be tested using Markov-chain models. However, in addition to the lab-based knowledge-push model, research instrumentalities provide another dynamics, for example, in medical fields (Price, 1984; Shinn, 2005; Von Hippel, 1976, 1988). Demand pull provides a third subdynamic, for example, in the case of need determination guided by the wish to cure diseases (Agarwal & Searls, 2008 and 2009).

In a recent study, Leydesdorff *et al.* (2012) suggested using visualizations to conceptualize the interactions among these three perspectives in terms of the different index structures of the Medical Subject Heading (MeSH) in the PubMed database, such as (*i*) "Diseases," (*ii*) "Drugs and Chemicals," and (*iii*) "Techniques and Equipment." As a projection in two dimensions, the three categories can be considered as continents of categories that provide a basemap for mapping trajectories in their mutual interactions. The non-linear dynamics of trajectories and regimes can be animated. As noted, the parameters of the statistical models are also expected to change in the case of path-dependent transitions. Using entropy statistics, however, one can begin to test these patterns and from there—using data-based bootstrapping techniques—it may be possible to specify confidence intervals as well.

**Conclusions and discussion**

The above is not intended as an exclusive plea in favor of using entropy statistics and/or MDS to study the dynamics of multivariate data. It seems noteworthy that the leading edge of scientometrics in the area of visualizations (e.g., Börner, 2010; Chen, 2004; Klavans & Boyack, 2011; van Eck & Waltman, 2010) has fruitfully profited from a research agenda that focuses on the nonlinear dynamics of science, technology, and innovation (Fagerberg *et al.*, 2005; Pyka & Scharnhorst, 2009; Scharnhorst *et al.*, 2012). The following scientometric research questions can provide a focus that require advancements in the statistics:

1. Dynamic models are different from static (or comparative static) ones; furthermore, dynamic models are different from evolutionary ones (e.g., Chen, 2006; Leydesdorff, 2010). Evolutionary dynamics contain feedback mechanisms and the dynamics may change because of bifurcations and path-dependent transitions;
2. One needs a transition to the domain of non-parametric statistics given the prevailing skewedness in scientometric distributions (Bornmann & Mutz, 2011; Leydesdorff *et al.*, 2011).

Advanced models in statistics such as vector autoregressions or crossed-lagged panel correlations have not yet been developed with this focus on the conceptual challenges in the scientometric domain. The number of variables included is limited and most models are hitherto developed for normally distributed data.[5]

Among the scientometric challenges, let me mention the desirable extensions of the study of the sciences as highly codified communication structures to translations at interfaces with technologies, and models of knowledge-based innovation as complex dynamics (Fagerberg *et al.*, 2005; Grilliches, 1994; Leydesdorff & Zawdie, 2010; Pyka & Scharnhorst, 2009). Both multivariate statistics and exploratory approaches which find their origins in engineering and physics can fruitfully be recombined for the exploration of the multi-perspectives approach advocated above (Helbing & Balietti, 2011; Leydesdorff, Rotolo, & De Nooy, in press). How can one conceptualize the recursive and interactive (sub)systems in terms of configurations so that this operationalization leads to results that can meaningfully be tested statistically?

---

[5] A different class of models is provided by actor-based models for network dynamics (e.g., Snijders *et al.*, 2010).